
\input phyzzx
\Pubnum{ MSUHEP--93/20 }
\date{March 1994}
\pubtype{ }
\def\journal#1&#2{#1#2 }

\def\ra{\rightarrow}
\def\d{{\rm d}}
\overfullrule=0pt
\def\joeb{1}
\def\zplota{2a}
\def\zplotb{2b}

\def\uacerna{3a}
\def\uacernb{3b}
\titlepage
\singlespace
\vskip3cm
\title{ The Nonperturbative Regime in QCD Resummation}
\author{ G.A. Ladinsky and C.--P. Yuan}
\address{Department of Physics and Astronomy,
Michigan State University \break
East Lansing, MI 48824 }
\vskip1cm
\abstract{
We study the nonperturbative functions in the Collins--Soper--Sterman
resummation formalism by examing Drell--Yan data in both
fixed target and collider experiments and then predict the transverse
momentum distributions of the $W^\pm$ and $Z^0$ bosons at the
Tevatron. Our results differ from that in the literature
and agree better with published CDF data.
Using statistical arguments, we find a $1\,$fb${}^{-1}$ luminosity
at the Fermilab Tevatron should be able to provide useful
constraints on the nonperturbative functions.
}
\vskip1cm
\vfill
\endpage
\chapter{Introduction}

\REF\whiggs{M. Swartz, {\it XVI International Symposium on Lepton--Photon
Interactions}, Cornell University, Ithaca, New York, 10-15 August 1993.}

The measurement of the mass ($M_W$) of the $W$ boson is very
important in testing the Standard Model (SM), which has
proven to be extremely successful.
This measurement is currently done at the Tevatron
by both the CDF and the D0 groups. In the framework of the SM,
the mass of the Higgs boson can be known to within a couple
of hundred GeV if $M_W$ is measured to within 100 MeV and the
mass of the top quark within 10 GeV.\refmark\whiggs

\REF\berger{E.L. Berger, {\it et al.}, Phys.~Rev. {\bf D40}, 83 (1989);
erratum {\it ibid}, {\bf D40}, 3789 (1989).}

Since the $W^\pm$ boson decays into a charged lepton and a neutrino, it is
important to know the kinematics of $W^\pm$ boson production for measuring
$M_W$, in particular, the transverse momentum distribution of the
$W^\pm$ bosons. The measurement of the transverse
momentum of the gauge bosons $V$ (either $W^\pm$, $Z^0$ or virtual
photon in the Drell--Yan process) also provides a test of QCD
in processes involving two--scale problems which are
beyond the usual framework of perturbative calculations in the
expansion of the strong coupling $\alpha_s$.  The measurement of
the rapidity asymmetry in the production of $W^\pm$ bosons provides
a handle on the $u/d$ ratio in parton distribution functions
such that quantities like $\Gamma_W/\Gamma_Z$ can be determined to
good precision.\refmark\berger

\REF\collins{J. Collins and D. Soper, Nucl. Phys. {\bf B193}, 381 (1981);
erratum {\it ibid}, {\bf B213}, 545 (1983);
{\it ibid}, {\bf B197}, 446 (1982).}
\REF\sterman{J. Collins, D. Soper and G. Sterman, Nucl. Phys. {\bf B250},
199 (1985).}

To obtain the kinematics of the gauge boson
we must resum the multiple soft and collinear gluon effects.
We adopt the Collins--Soper--Sterman (CSS) formalism\refmark{\collins,\sterman}
to resum these multiple gluon effects and closely
follow the notation used in Ref.~\sterman.
In this paper we study
the nonperturbative part of this resummation
formalism and show the importance of its contribution to the
transverse momentum distribution of the gauge bosons.

\REF\fnotea{Note Eq.~\twosixone\ represents the
nonperturbative factor differently than Ref.~\sterman, and
we have incorporated the flavor dependent couplings into
$\widetilde{W}_j$.}

To describe the kinematics of the gauge boson for all transverse momenta
$Q_T$, we use the resummed formula differential in the mass ($\sqrt{Q^2}$),
rapidity ($y$), and $Q_T$ of the boson $V$,\refmark{\sterman}
$$\eqalign{
{ \d \sigma(AB \ra V) \over \d Q^2 \d y \d Q^2_T} & =
{\pi \over S} \sigma_0 \, \delta(Q^2-M^2_V)
\bigg\{ {1\over (2 \pi)^2}
\int_{}^{} \d^2 b \, e^{i {\vec Q_T} \cdot {\vec b}} \,
\sum_j
\widetilde{W}_j (b_*;Q,x_A,x_B) \cr
& \times \exp \left[-\ln\left(Q\over 2Q_0\right)h^{}_Q(b)
-h_{j/A}(x_A,b)-h_{\bar{j}/B}(x_B,b)\right] \cr
& + Y(Q_T;Q,x_A,x_B) \bigg\}, \cr}
\eqn\twosixone
$$
where $\sigma_0$ provides the process--dependent normalization and
$\widetilde{W}$ and $Y$ are derived from pertrubative
calculations.\refmark{\fnotea}


The functions $h_{j/A}$ and $h_{\bar{j}/B}$,
which carry a flavor dependence as well as their respective dependence
on the momentum fractions $x_A$ and $x_B$,
handle the nonperturbative behaviour at large $b$
along with $h^{}_Q$. These functions
are to be obtained by a fit to data, subject to the constraint that they
must vanish when $b\ra 0$. The constant $Q_0$, in Eq.~\twosixone,
is completely arbitrary.

\REF\davies{C. Davies, Ph.D. Thesis, Churchill College (1984);
C.~Davies and W.~Stirling, Nucl. Phys. {\bf B244} (1984) 337;
C.~Davies, B.~Webber and W.~Stirling, Nucl. Phys. {\bf B256} (1985) 413.}
\REF\arnold{P.B.~Arnold and R.P.~Kauffman, Nucl.~Phys. {\bf B339} (1991) 381.}
\REF\duke{D.W.~Duke and J.F.~Owens, Phys. Rev. {\bf D30} (1984) 49.}
\REF\ito{A.S. Ito, et al., Phys. Rev. {\bf D23} (1981) 604.}
\REF\joe{J. Paradiso, Ph.D. Thesis,
Massachusetts Institute of Technology (1981);\hfill\break
D. Antreasyan, et al., Phys. Rev. Lett. {\bf 47} (1981) 12.}

Among the nonperturbative functions used previously
in the CSS formalism
are those of Davies, Stirling and Webber.\refmark{\davies,\arnold}
They selected the functional form that provided a gaussian smearing of
the transverse momentum,
$$
h^{}_Q(b)=\tilde{g}_2b^2
\qquad
\hbox{and}
\qquad
h_{j/A}(x_A,b)+h_{\bar{j}/B}(x_B,b)=\tilde{g}_1b^2,
\eqn\oldform
$$
where $\tilde{g}_1,\tilde{g}_2$ are phenomenological constants.
For simplicity
the nonperturbative functions are assumed to be independent of flavor
and momentum fractions $x_A$, $x_B$.
Using the parton distribution functions of Duke and Owens,\refmark{\duke}
their favored
values after fitting to E288\refmark{\ito}
and R209\refmark{\joe}
Drell--Yan data
with $Q_0=2\,$GeV and $b_{max}=0.5\,\hbox{GeV}^{-1}$ were
$$
\tilde{g}_1=0.15\ \hbox{GeV}^{2}
\qquad
\hbox{and}
\qquad
\tilde{g}_2=0.4\ \hbox{GeV}^{2}.
\eqn\oldset
$$
The interest at that time was the production of $W^\pm$ bosons
in hadron collisons at $\sqrt{s}=540\,$GeV.

\REF\cteq{J. Botts, J.~Huston, H. Lai, J.~Morfin, J.~Owens,
J.~Qiu, W.--K.~Tung, H.~Weerts, Michigan State University
preprint MSUTH-93/17.}
\REF\hmrs{P.N. Harriman, A.D. Martin, R.G.~Roberts and W.J.~Stirling,
\hfill\break
Phys.~Rev.~{\bf D42}~(1990)~798.}
\REF\wcdf{F. Abe, et al., Phys. Rev. Lett. {\bf 66} (1991) 2951;
B.L.~Winer, Ph.D. Thesis (1991),
Lawrence Berkeley Laboratory preprint LBL-30221.}
\REF\zcdf{F. Abe, et al., Phys. Rev. Lett. {\bf 67} (1991) 2937;
J.S.T.~Ng, Ph.D. Thesis (1991), Harvard University preprint HUHEPL-12.}

\chapter{A New Study}

For this study we chose $Q_0=1.6\,$GeV and
$b_{max}=0.5\,\hbox{GeV}^{-1}$ with a different functional form for the
nonperturbative part,
$$
h^{}_Q(b)=g_2b^2
\qquad
\hbox{and}
\qquad
h_{j/A}(x_A,b)+h_{\bar{j}/B}(x_B,b)=g_1b[b+g_3\ln(100x_Ax_B)],
\eqn\newform
$$
where $g_1,g_2,g_3$ are phenomenological constants.
Since we want to investigate the
production of gauge bosons at the Fermilab Tevatron
with $\sqrt{s}=1.8\,$TeV, our kinematic region includes values of
$\tau=x_Ax_B$ that are significantly lower than those relevant to
Ref.~\davies.  For this reason we refit for the
nonperturbative functions using lower mass ranges
in the R209
($pp\rightarrow \mu^+\mu^-+X$ at $\sqrt{s}=62\,$GeV)
and E288
($pN\rightarrow \mu^+\mu^-+X$ at $\sqrt{s}=27.4\,$GeV)
data in addition to the data from CDF
($p\bar{p}$ collisions at $\sqrt{s}=1.8\,$TeV)
for $Z^0$ production.
These lower $\tau$ values used from the R209 and E288 data, however,
are still roughly a factor of
two larger than the typical $\tau$ value of $W^\pm$ and $Z^0$ physics
at the Tevatron.  To improve these nonperturbative functions in the
CSS formalism, we postulate the $x_A,x_B$ dependence for the nonperturbative
functions $h_{j/A}$ and $h_{\bar{j}/B}$ as in Eq.~\newform.
Our choice of $\ln(x_Ax_B)$ is inspired by the fact that the average
transverse momentum of the Drell--Yan pair grows
slowly with $\tau$.\refmark{\ito}
When better statistics becomes available,
these nonperturbative functional forms also can be extracted
for the lower $\tau$ kinematics
through the study of Drell--Yan pairs and $Z^0$ boson production
at the Tevatron collider.

For $W^\pm$ boson studies at the Tevatron, it would be excellent if one had
very high statistics for the $Z^0$ data, since this
would enable a determination of
the nonperturbative functions in a kinematic region that overlaps greatly
with the kinematics for $W^\pm$ boson production, but such high statistics
data do not exist.
Using the R209 data over the mass range $5<Q<8\,$GeV in conjucntion with
the $Z^0$ boson data from CDF, $g_2$ was determined.
We note that $g_2$ is associated with the $\ln({Q\over 2Q_0})$ factor in the
nonperturbative functions.
With that $g_2$, the
same R209 data were taken in combination with the E288 data over
$6<Q<8\,$GeV  to obtain an $x_A,x_B$ dependence by determining $g_1$ and $g_3$.
Using the CTEQ2M parton distribution functions (PDFs)\refmark{\cteq}
with $Q_0=1.6\,$GeV and $b_{max}=0.5\,\hbox{GeV}^{-1}$, the nonperturbative
parameters are
$$
{g}_1=0.11^{+0.04}_{-0.03}\ \hbox{GeV}^{2},
\qquad\qquad
{g}_2=0.58^{+0.1}_{-0.2}\ \hbox{GeV}^{2},
\qquad\qquad
{g}_3=-1.5^{+0.1}_{-0.1}\ \hbox{GeV}^{-1}.
\eqn\newset
$$

\REF\uaref{UA2 Collaboration (J.~Alitti et al.),
Z. Phys. {\bf C47} (1990) 523.}
\REF\hera{H1 Collaboration (I. Abt, et al.), DESY-93-117, Aug 1993;
ZEUS Collaboration (M. Derrick, et al.), Phys.~Lett.~{\bf B316}~(1993)~412}
\REF\allcteq{R. Brock, G. Ladinsky, W.--K. Tung, C.--P. Yuan, in preparation.}

We now proceed to discuss the comparison of the $Q_T$ distributions
obtained from the previous values of $\tilde{g}_1,\tilde{g}_2$ with
that given by $g_1,g_2,g_3$.
For the high $\tau$ kinematics of
the R209 data with $11<Q<25\,$GeV, the $d\sigma /dQ_T^2$
distribution given by the nonperturbative form of Eq.~\oldset,
where we naivley carry over the $\tilde{g}$
to the CTEQ2M PDF,
lies very close to the results provided by using the CTEQ2M PDF with
Eq.~\newset.
The results provided by Eq.~\newset\ also are consistent with
the E288 Drell--Yan data, whose typical $\tau$ values range
from about $1\over 5$ to $1\over 2$.
For the $5<Q<8\,$GeV range in the R209 data, where
the $\tau$ values probed are much smaller and more relevant to
$W^\pm$ and $Z^0$ boson physics at the Fermilab Tevatron,
Fig.~\joeb\ shows that the
old form of Eq.~\oldset\ with the CTEQ2M PDF is quite different from the
result of Eq.~\newset, deviating from the experimental result.  This
is a clear demonstration that for precision measurements (like the
determination of $M_W$ or $\Gamma_W$) or simply for theoretical consistency,
it is necessary,
particularly when entering a new kinematic region
or changing the PDF used,
to have a nonperturbative parametrization
consistent with that PDF and kinematics.
In Fig.~\joeb\ we also show
the result obtained when the favored values of Ref.~\davies\ are used
in conjuction with the HMRSB PDF,\refmark{\hmrs}  which has been used in
previous publications.
This too does not agree with the data,
clearly discriminating between the two nonperturbative functions of
Eq.~\oldset\ and Eq.~\newset.
The former parametrization is invalid in this kinematic region
independent of the PDF used.

Having a better nonperturbative dependence for describing the low $Q_T$
kinematics of the $W^\pm,Z^0$ boson production at the Fermilab Tevatron,
we can see if there is any change from previous expectations.
In Fig.~\zplota\ (\zplotb)
we display three calculations
against the results for $Z^0$ ($W^\pm$) production
obtained by the CDF collaboration.
(We use $M_W=80\,$GeV and $M_Z=91.17\,$GeV.)
The dashed (dashed-dotted) curve is where we took the previous values of
Eq.~\oldset\ with the CTEQ2M (HMRSB) PDF, while the solid curve represents
the results of the new fit with the CTEQ2M PDF.  Simply comparing
the new set with the old set as applied to the CTEQ2M PDF, it is
apparent that the peak of the $Q_T$ spectrum has dropped and shifted
to a higher $Q_T$ value.  The HMRSB result has a lower
peak height, but the peak remains at the lower $Q_T$, just as the dashed curve.
It is not due to the comparison of these theoretical results with the
low statistics data that favor one result over the other, rather, it
is the inability of the Eq.~\oldset\ to account for the Drell-Yan results
with the more pertinent $\tau$ kinematics that decides.
Recall that at very low $Q_T$ the
$d\sigma /dQ_T$ experimental results\refmark{\wcdf} for the $W^\pm$ boson were
below theoretical expectations with the old fits
and that as the $Q_T$ rose the data quickly gave
experimental results above theoretical expectations.\refmark{\zcdf}
The new fit not only shifts the peak in the $d\sigma /dQ_T$ distribution to
higher $Q_T$, it also makes the peak broader, thereby yielding an improvement
between the data and the theory in the region of $5\le Q_T\le 20\,$GeV.
In Fig.~\uacerna\ (\uacernb)
we show the $Q_T$ distribution at $y=0$ for the $Z^0$ ($W^\pm$) boson
relevant to the CERN collider.\refmark{\uaref}

\REF\mrsdmp{A.D. Martin, W.J. Stirling, and R.G. Roberts,
Phys.~Lett.~{\bf B306} (1993) 145; \break
erratum, {\it ibid}, {\bf B309} (1993) 492.}

We observe that the results with the old parameters are similar
between the CTEQ2M and HMRSB PDFs for the $\tau$ region of Fig.~\joeb,
yet as we enter the lower $\tau$ kinematics of the CDF data, the
difference appears to grow.  In this regard, we note that the CTEQ2M
PDF fits with the low--$x$ $F_2$ data from HERA,\refmark{\hera} while
HMRSB does not.  Using the new nonperturbative parameters with the
MRSD$-^\prime$ PDFs\refmark{\mrsdmp} gave results that differed only negligibly
when compared against the CTEQ2M results for $Q_T<20\,$GeV.

\chapter{Measuring the Nonperturbative functions at the Fermilab Tevatron}

\REF\sterman{G. Sterman, private communication.}

\REF\minuit{MINUIT, Application Software Group, Computing and Networks
Division, CERN, Geneva, Switzerland.}

Measuring the nonperturbative functions
in the CSS resummation formalism provides information about the
nonperturbative nature of QCD theory. For instance,
$h^{}_Q(b)$ in Eq.~\oldform\
might have an operator definition as the vacuum expectation of the
gluon condensate.\refmark\sterman

We explore the necessary luminosity and energy of the
upgraded Tevatron to perform this kind of study.
To isolate the problem interested, we assume that the parton
distribution functions are known in the relevant kinematic
region for producing $W^\pm$ and $Z^0$ bosons at the Tevatron.
In Fig.~4  we show the transverse momentum distributions
for the $W^\pm$ and $Z^0$ bosons produced at a $p\bar p$ collider
with $\sqrt{s}=1.8,3.5\,$TeV
using the CTEQ2M PDFs.

For this study we chose $Q_0=1.6\,$GeV,
fixed $g_3=-1.5\,\hbox{GeV}^{-1}$
and pretended that
nature wants $g_1=0.11\,\hbox{GeV}^2$ and $g_2=0.58\,\hbox{GeV}^2$.
To estimate the accuracy for the measurement of the nonperturbative
parameters as a function of luminosity,
fake data was generated for the
$d\sigma/dp_T$ distribution describing the production of $Z^0$ bosons,
where a branching fraction of $0.06$ was included to focus only on
the $Z\rightarrow e^-e^+$ and $Z\rightarrow \mu^-\mu^+$ decay
channels.
This fake data was then fit to theory using MINUIT\refmark\minuit to both
determine the best values for the parameters $g_1$ and $g_2$ and to
estimate the errors.

To create the fake data for the $d\sigma/dp_T$ distribution, a bin width
of $\Delta p_T=1\,$GeV was assumed.
Given the theoretical value for $d\sigma/dp_T$ and the luminosity ($\cal L$),
a statistical error was evaluated for each value of $d\sigma/dp_T$,
$$
\varepsilon_{stat}=\sqrt{d\sigma/dp_T\over \Delta p_T {\cal L}}\ .
\eqn\staterr
$$
Assigning a detector uncertainty equivalent to the statistical error for the
purposes of this estimate, the error for each of the fake data points in
$d\sigma/dp_T$ becomes
$$
\varepsilon=\sqrt{2}\,\varepsilon_{stat}
=\sqrt{2d\sigma/dp_T\over \Delta p_T {\cal L}}\ .
\eqn\allerr
$$
Taking eight values at $p_T=1,2,\ldots,8\,$GeV, each point of the
$d\sigma/dp_T$ theory was
randomized according to a gaussian distribution about its theoretical
value using a width of $\varepsilon$.  This provided a fake data sample
to use for the fit of the nonperturbative functions.

When the fake data was fit to
the theoretical values, the errors provided by MINUIT demonstrated
that $g_1$ can be known to within about $6\%$ ($20\%,50\%$) and $g_2$
can be known to within about $2\%$ ($4\%,15\%$) at the $95\%$
confidence level given a luminosity of $10\,\hbox{fb}^{-1}$
($1\,\hbox{fb}^{-1}$,$0.1\,\hbox{fb}^{-1}$).

The accuracy for the measurement of the nonperturbative
functions in the CSS resummation formalism is relatively indifferent
to the energy upgrade of the Tevatron
(within a factor of two for $1.8\,$TeV and $3.5\,$TeV energies)
when compared against an orders of magnitude increase
in the luminosity,
indicating that the luminosity of the machine is crucial.
In this analysis, we only assume $Z \ra e^+e^-$ and
$Z \ra \mu^+\mu^-$ data. If the mass of the $W$ boson is known, one
can include the data from $W^\pm \ra l^\pm+\hbox{neutrino}$
(for $l=e$ or $\mu$)
to further improve the measurement by about a factor of two.

Although this study is merely a theoretical exercise,
the nonperturbative
functional forms as described in Eq.~\newform\ might need to be
revised when better data become available.  It eventually may prove
that the flavor dependence cannot be ignored.
Nevertheless, our point is that data from Drell--Yan experiments
and the production of
$W^\pm$ and $Z^0$ bosons at hadron colliders and fixed target experiments
like those at Fermilab will be instrumental in providing bounds on the
nonperturbative structure at low $Q_T$.

\chapter{Conclusion}

For precision measurements (like the
determination of $M_W$, $\Gamma_W$, or the charge asymmetries
for $W^\pm$ bosons)
or simply for theoretical consistency,
it is necessary,
particularly when entering a new kinematic region
or changing the PDF used,
to have a nonperturbative parametrization
consistent with that PDF and kinematics.
In this study we have demonstrated
that in the CSS resummation formalism
represented by Eq.\twosixone, the contributions
of the nonperturbative functions to the $Q_T$
distribution of Drell--Yan pairs and the production
of $W^\pm$ and $Z^0$ bosons at fixed target and
hadron colliders
are important.  When the physics of interest ({\it e.g.}, low $Q_T$
boson production) probes different kinematics, such as lower
momentum fractions $x_A$ and $x_B$,
these nonperturbative functions have to be reevaluated, just as the
PDFs have to be updated when new data probing smaller $x$ regions
become available.
Consistency between the data from
Drell--Yan processes and the production of
$W^\pm$ and $Z^0$ bosons will not only support QCD theory, but also will
provide a tool which can
facilitate our pursuits in physics beyond QCD through a better
understanding of signal and background processes in this important
kinematic region where event rates are large.
As an example, we have shown that the theoretical
$Q_T$ distributions for $W^\pm$
and $Z^0$ bosons at the Fermilab Tevatron
agree better with
experimental data after a new fit for the nonperturbative dependence
has been performed using Drell--Yan data proximate to the
relevant kinematic regions in $\tau=x_Ax_B$
for $W^\pm$ and $Z^0$ boson production at
the Fermilab Tevatron.

For the precision measurement of $M_W$ and testing QCD theory in processes
involving two--scale physics, such as the $W^\pm$, $Z^0$ and Drell--Yan pair
production, it is important to know the theoretical errors due to the
nonperturbative parametrization, the factorization scale dependence,
and the parton distribution functions.  All of these considerations are
under study.\refmark{\allcteq}

\vfill\eject

\leftline{\bf Acknowledgements}

\vskip 0.5cm

We thank our colleagues P.~Agrawal, J.~Botts, G.~Brandenburg,
R.~Brock, J.~Collins, K.~Einsweiler, J.~Huston, H.~Lai, J.~Ng,
J.~Paradiso, B.~Pope, J.~Qiu, D.~Soper, G.~Sterman, W.--K.~Tung,
H.~Weerts, and B.~Winer.  This work was performed in part under TNRLC
grant RGFY9240 and NSF grant PHY93-09902.

\vfill
\endpage
\refout
\vfill
\endpage
\centerline{\fourteenpoint Figure Captions}
\vskip 0.5cm
\item{1.}{Comparison of R209 Drell--Yan data
with calculations using Eq.~\newset\
and the CTEQ2M PDF or Eq.~\oldset\ using either the CTEQ2M or HMRSB PDFs.}
\item{2.}{Comparison of (a) $Z^0$ boson or (b) $W (=W^++W^-)$ boson
production at CDF with calculations using Eq.~\newset\
and the CTEQ2M PDF or Eq.~\oldset\ using either the CTEQ2M or HMRSB PDFs.}
\item{3.}{Calculations as in Fig.~2, except for $\sqrt{s}=630\,$GeV.}
\item{4.}{ The transverse momentum distribution for the production of
(a)~Z and (b)~W bosons in $p\bar{p}$ collisions at $\sqrt{s}=1.8,3.5\,$TeV.}

\vfill\eject

\vfill
\endpage





\end
\bye